# Quantum cards and quantum rods

M. Batista[1] and J. Peternelj[2]

[1] Faculty of Maritime Studies and Transport, University of Ljubljana, 6320 Portorož, Pot pomorščakov 4, Slovenia

[2] Division of Mathematics and Physics, Faculty of Civil and Geodetic Engineering, University of Ljubljana, 1001 Ljubljana, Jamova 2, Slovenia

Quantum mechanical analysis of a rigid rod with one end fixed to a flat table is presented. It is shown, that for a macroscopic rod the ground state is orientationally delocalized only if the table is absolutely horizontal. In this latter case the rod, assumed to be initally in the upright orientation, falls down symmetrically and simultaneously in both directions, as claimed by Tegmark and Wheeler. In addition, the time of fall is calculated using WKB wavefunctions representing energy eigenstates near the barrier summit.

## I. INTRODUCTION

In their historical review »100 Years of Quantum Mysteries«[1] Tegmark and Wheeler discussed the following problem. »......You take a card with a perfectly sharp edge and balance it on its edge on a table. According to classical physics, it will in principle stay balanced forever. According to the Schrödinger equation, the card will fall down in a few seconds even if you do the best possible job of balancing it, and it will fall down in both directions-to the left and to the right-in superposition.«

In what follows we shall, instead of a card, consider the rotational motion of a thin rigid rod with one end fixed to a horizontal table. If the length of the rod and its mass are denoted as $l$ and $m$, respectively, then its energy can be written as $E = \frac{1}{2}J\omega^2 + V_0\cos\theta$. $J = ml^2/3$, is the moment of inertia of the rod with respect to the fixed end, $V_0 = mgl/2$, $\theta$ is the angle of the rod with respect to the vertical, $\omega = d\theta/dt$, is the angular velocity of the rod and $g$ is the acceleration of gravity. Let us assume that the rod is released from rest from an initial orientation $\delta\theta << 1$. The time of fall $t_{class.}$ is obtained from the law of conservation of energy and is given as[2],

$$t_{class.} = \int_{\delta\theta}^{\pi/2} \frac{d\theta}{\omega} = \sqrt{\frac{l}{3g}} \int_{\delta\theta}^{\pi/2} \frac{d\theta}{\sqrt{(\cos\delta\theta - \cos\theta)}} \cong \frac{1}{\omega_c} \{\ln[8(\sqrt{2}-1)] - \ln\delta\theta\}, \qquad (1.1)$$

where $\omega_c = \sqrt{V_0/J} = \sqrt{3g/2l}$ is the classical oscillation frequency of the rod pendulum. Therefore, $t_{class.} \to \infty$, when $\delta\theta \to 0$. That this is not the case when the rod is treated quantum mechanically, could be suspected by considering the uncertainty principle for the angular momentum of the rod, $L = J\omega$, and the angle $\theta$. If the orientation of the rod is determined to be within the range $\delta\theta$ around the vertical position, then the uncertainty in its angular momentum is, $\delta L \cong \hbar/\delta\theta$, and the initial uncertainty in the angular velocity is $\delta\omega = \delta L/J \cong \hbar/J\delta\theta$. Consequently, the rod will inevitably start falling with some nonzero angular velocity which, moreover, is increasing with decreasing $\delta\theta$. Thus we anticipate that in quantum mechanics the $\ln\delta\theta$ term in (1.1) must be somehow prevented from becoming too large.

In the next section we will consider the quantum mechanical analysis of the rotational motion of a rigid rod with one end fixed to a horizontal table. No really new results will emerge from this calculation which represents merely an application of well known formulas to a specific and rather simple problem which is, nevertheless, conceptually quite interesting.

## II.   THE QUANTUM ROD

The classical Hamiltonian of the rod rotating in the vertical plane with one end fixed (Fig.1) is, $H = L^2/2J + V(\theta)$. Assuming a rigid rod and a rigid horizontal table, the potential energy is (Fig.2),

$$V(\theta) = \begin{cases} V_0 \cos\theta, & -\frac{\pi}{2} \leq \theta \leq \frac{\pi}{2} \\ \infty, & |\theta| > \frac{\pi}{2} \end{cases}. \qquad (2.1)$$

Transition to quantum mechanics is achieved by writing, $L = -i\hbar d/d\theta$, which yields the Hamiltonian

$$H = -\frac{\hbar^2}{2J}\frac{d^2}{d\theta^2} + V(\theta). \qquad (2.2)$$

The energy eigenfunctions $\psi(\theta)$ are determined as solutions of the time-independent Schrödinger equation,

$$-\frac{\hbar^2}{2J}\frac{d^2\psi}{d\theta^2} + V_0\cos\theta\,\psi = E\psi, \tag{2.3}$$

subject to the boundary conditions $\psi(\theta = \pm\pi/2) = 0$. Since the Hamiltonian (2.2) is invariant with respect to the transformation $\theta \to -\theta$, we can classify the eigenfunctions as symmetric (even) $\psi^{(+)}$ and antisymmetric (odd) $\psi^{(-)}$ solutions of (2.3), such that $\psi^{(\pm)}(-\theta) = \pm\psi^{(\pm)}(\theta)$. The corresponding eigenvalues will be denoted as $E^{(\pm)}$.

As seen from Fig.2, the potential $V(\theta)$ consists of two wells separated by a barrier. For energies $E \ll V_0$, tunneling through the barrier is not appreciable and the motion of the rod is confined predominantly to one well or the other. Let us denote the corresponding wavefunctions and the associated energies as $\psi_0(\pm\theta)$ and $E_0$, respectively. To account for tunneling we construct approximate solutions of (2.3) as,

$$\psi^{(\pm)}(\theta) = \frac{1}{\sqrt{2}}[\psi_0(\theta) \pm \psi_0(-\theta)], \tag{2.4}$$

and the energies $E^{(\pm)}$ are,

$$E^{(\pm)} = E_0 \mp \Delta E, \tag{2.5}$$

where $2\Delta E$ represents the tunneling splitting of the level $E_0$ which is doubly degenerate in the absence of tunneling. Using (2.4), (2.5) and the time-dependent Schrödinger equation $i\hbar\partial\psi/\partial t = H\psi$, one can show that $2\Delta E/\hbar$ represents the rate of tunneling from one well to the other, if the rod was initially localized in one of the wells. On the other hand, it is natural to expect that the rate of tunneling is proportional to the number of times per second (attempt frequency) that the rod, with energy $E_0 < V_0$, approaches the classical turning point , multiplied by the WKB amplitude for barrier penetration,

$$e^{-W_0} = e^{-\frac{1}{\hbar}\int_{-\theta_0}^{\theta_0} d\theta \sqrt{|2J(E_0 - V_0 \cos\theta)|}}, \qquad (2.6)$$

where $\theta_0 = \arccos(E_0/V_0)$. If $T$ is the period of classical motion of the rod in either one of the wells we can write,

$$\Delta E \propto \frac{\hbar\omega}{\pi} e^{-W_0}, \qquad (2.7)$$

and $\omega = 2\pi/T$.

From (2.4) it is seen that the low energy eigenstates and, in particular, the ground state of the rod are not orientational eigenstates, but are instead a linear superposition of states, such that the rod seems to be simultaneously to the left and to the right of the fixed end, in agreement with Tegmark and Wheeler[1]. Furthermore, it is already clear from (2.7) that energy levels below the barrier are paired into doublets, formed by pairs of neighboring states of opposite parity. The pairing effect, does not, however, vanish at the top of the barrier as one would perhaps expect from (2.7), but goes over smoothly towards the spectrum of a »particle« in an infinite potential well, resulting from (2.3), in the limit $E \to \infty$. We can verify this rather easily by numerical solution of (2.3) which can be transformed into standard Mathieu form[3] by introducing a new variable $\theta = 2\eta$,

$$\frac{d^2\psi}{d\eta^2} + (a - 2q\cos 2\eta)\psi = 0, \qquad (2.8)$$

where, $a/4 = E/(\hbar^2/2J)$ and $q/2 = V_0/(\hbar^2/2J)$, while the boundary conditions are $\psi(\pm\pi/4) = 0$. The solution of (2.8) can be written in general as,

$$\psi(\eta) = C_1 Ce(a,q,\eta) + C_2 Se(a,q,\eta), \qquad (2.9)$$

where $Ce(a,q,\eta)$ and $Se(a,q,\eta)$ are, respectively, even and odd general Mathieu functions. Imposing the boundary conditions supplies us with two equations,

$$C_1 Ce(a,q,\pi/4) \pm C_2 Se(a,q,\pi/4) = 0. \qquad (2.10)$$

To obtain a nontrivial solution, the determinant of the above system of equations must be zero, i.e.,

$$Ce(a,q,\pi/4)Se(a,q,\pi/4) = 0. \tag{2.11}$$

Consequently, the solutions of (2.8) are,

$$\psi_n^{(\pm)}(\eta) = \begin{cases} C_n^{(+)}Ce(a_n^{(+)},q,\eta) \\ C_n^{(-)}Se(a_n^{(-)},q,\eta) \end{cases}, \tag{2.12}$$

where $n$ is a positive integer. The characteristic values $a_n^{(\pm)}$ are determined from, $Ce(a_n^{(+)},q,\pi/4) = 0$, and $Se(a_n^{(-)},q,\pi/4) = 0$, while the constants $C_n^{(\pm)}$ are fixed by the normalization condition. The corresponding energy eigenvalues are,

$$E_n^{(\pm)} = \frac{\hbar^2}{2J} \cdot \frac{1}{4} a_n^{(\pm)}. \tag{2.13}$$

Some of the results are presented in Table 1 and Fig. 3.

It has been said[1] that the quantum card, initially balanced in the upright orientation, will fall down in both directions at once. The card's inital wavefunction, representing the card in the vertical orientation, changes continuously through a series of states such that at any time the card appears to occupy two opposite orientations ($\pm\theta$) at once. Of course, the same is true also for the motion of the quantum rod, considered here.

Let us take the initial state of the rod in the upright orientation to be ($\psi(\theta,t=0) \equiv \psi(\theta,0)$),

$$\psi(\theta,0) = \pi^{-1/4}\sigma^{-1/2}e^{-\theta^2/2\sigma^2}. \tag{2.14}$$

The uncertainty of the initial orientation of the rod corresponding to this state, as defined by the standard deviation[4], is $\delta\theta = \sigma/\sqrt{2}$, while the uncertainty in its angular momentum is $\delta L = \hbar/\sigma\sqrt{2}$. Consequently, $\delta\theta\delta L = \hbar/2$, in agreement with the uncertainty principle mentioned earlier.

The wavefunction at some later time $t$ is obtained from the time dependent Schrödinger equation as,

$$\psi(\theta,t) = e^{-\frac{i}{\hbar}Ht}\psi(\theta,0) = \sum_n c_n^{(+)} e^{-\frac{i}{\hbar}E_n^{(+)}t}\psi_n^{(+)}(\theta), \qquad (2.15)$$

where,

$$c_n^{(+)} = \int_{-\pi/2}^{+\pi/2} d\theta \psi_n^{(+)}(\theta)\psi(\theta,0), \qquad (2.16)$$

and the subscript $n$ was added to label different energy eigenstates of even parity. Since the Hamiltonian and the initial state (2.14) are both symmetric with respect to $\theta$, it is clear that $\psi(\theta,t)$ is also symmetric at all times and, consequently, the rod is falling in both directions at once. Furthermore, during time evolution the expectation value of energy of the rod is constant and is equal to,

$$E = -\frac{\hbar^2}{2J}\int_{-\infty}^{+\infty}\psi(\theta,0)\frac{d^2\psi(\theta,0)}{d\theta^2}d\theta + V_0\int_{-\infty}^{+\infty}\cos\theta\,\psi^2(\theta,0)d\theta. \qquad (2.17)$$

Using (2.14), we obtain,

$$E = \frac{\hbar^2}{2J}\frac{1}{2\sigma^2} + V_0 e^{-\sigma^2/4} \cong \frac{\hbar^2}{2J}\frac{1}{2\sigma^2} + V_0\cos\left(\frac{\sigma}{\sqrt{2}}\right) \cong V_0\left[1 - \frac{\sigma^2}{4} + \frac{\hbar^2/2J}{V_0}\frac{1}{2\sigma^2}\right]. \qquad (2.18)$$

Requiring that the expectation value of energy of the quantum rod in the upright orientation is roughly equal to $V_0$, the energy of the classical rod in vertical position, constrains the width of the initial state $1 \gg \sigma \gg (\hbar/J\omega_c)^2$. In particular, if we choose $\sigma = (\hbar/J\omega_c)^{1/2}$, we obtain $E = V_0$ Now, recalling (1.1), we tentatively suggest that the time of fall for the quantum rod, whose initial state is (2.14) with the width $\sigma = (\hbar/J\omega_c)^{1/2}$, is

$$t'_Q = \frac{1}{\omega_c}\left\{\ln[8(\sqrt{2}-1)] - \ln\left(\frac{\hbar}{J\omega_c}\right)^{1/2}\right\}. \tag{2.19}$$

For a 10 cm long rod with mass equal $10^{-3}$kg, (2.19) yields $t'_Q \cong 3$sec.

All the results presented above will be confirmed in the next section using WKB approximation.

### III. WKB APPROXIMATION

Let us calculate the eigenstates $\psi^{(\pm)}(\theta)$ and the corresponding eigenvalues $E^{(\pm)}$ using WKB approximation[4]. The solutions of the Schrödinger equation, corresponding to energies $E < V_0$ ($E$ should not be too close to the barrier summit), can be written in the classically inaccessible region, $-\theta_0 < \theta < \theta_0$, as

$$\psi(\theta) = \frac{C_1}{\sqrt{|L|}} e^{-\frac{1}{\hbar}\int_\theta^{\theta_0} d\theta|L|} + \frac{C_2}{\sqrt{|L|}} e^{\frac{1}{\hbar}\int_\theta^{\theta_0} d\theta|L|}, \tag{3.1}$$

where, $L = \sqrt{2J(E - V_0\cos\theta)}$, is the classical angular momentum of the rod with respect to the fixed end and, $\theta_0 = \arccos(E/V_0)$, is the classical turning point. If we now require that the wavefunctions be even or odd, that is, $\psi^{(\pm)}(-\theta) = \pm\psi^{(\pm)}(\theta)$, we obtain $C_2 = \pm C_1 e^{-W}$ and, consequently,

$$\psi^{(\pm)}(\theta) = \frac{C^{(\pm)}}{\sqrt{|L|}}\left(e^{-\frac{1}{\hbar}\int_\theta^{\theta_0} d\theta|L|} \pm e^{-W} e^{\frac{1}{\hbar}\int_\theta^{\theta_0} d\theta|L|}\right). \tag{3.2}$$

where, $W = \int_{-\theta_0}^{\theta_0} d\theta|L|/\hbar$, is analogous to $W_0$.

In the classically allowed region, say $\theta > \theta_0$, the WKB wavefunctions have the form,

$$\psi(\theta) = \frac{C'}{\sqrt{L}} \cos\left\{ \frac{1}{\hbar}\int_{\theta_0}^{\theta} d\theta L - \frac{\pi}{4} + \delta \right\} = \frac{C'\cos\delta}{\sqrt{L}}\cos\left( \frac{1}{\hbar}\int_{\theta_0}^{\theta} d\theta L - \frac{\pi}{4} \right) - \frac{C'\sin\delta}{\sqrt{L}}\sin\left( \frac{1}{\hbar}\int_{\theta_0}^{\theta} d\theta L - \frac{\pi}{4} \right).$$

(3.3)

To establish the correspondence between (3.2) and (3.3), we employ the connection formulas[4,5,6],

$$\frac{1}{\sqrt{|L|}} e^{-\frac{1}{\hbar}\int_{\theta}^{\theta_0} d\theta |L|} \rightarrow \frac{2}{\sqrt{L}}\cos\left( \frac{1}{\hbar}\int_{\theta_0}^{\theta} d\theta L - \frac{\pi}{4} \right), \text{ and } \frac{1}{\sqrt{L}}\sin\left( \frac{1}{\hbar}\int_{\theta_0}^{\theta} d\theta L - \frac{\pi}{4} \right) \rightarrow -\frac{1}{\sqrt{|L|}} e^{\frac{1}{\hbar}\int_{\theta}^{\theta_0} d\theta |L|},$$

which yield,

$$\psi^{(\pm)}(\theta > \theta_0) = \frac{2C^{(\pm)}\left[ 1 + \left( \frac{1}{2}e^{-W} \right)^2 \right]^{1/2}}{\sqrt{L}} \cos\left\{ \frac{1}{\hbar}\int_{\theta_0}^{\theta} d\theta L - \frac{\pi}{4} \pm \arctan\left( \frac{1}{2}e^{-W} \right) \right\}. \quad (3.4)$$

The form of the WKB wavefunctions in the left well, $\psi^{(\pm)}(\theta < -\theta_0)$ is, by symmetry, equal to $\pm\psi^{(\pm)}(\theta > \theta_0)$. Imposing the boundary condition, $\psi^{(\pm)}(\theta = \pi/2) = 0$, provides us with the WKB quantization condition for energy,

$$\frac{1}{\hbar}\int_{\theta_0}^{\pi/2} d\theta\sqrt{2J(E - V_0\cos\theta)} - \frac{\pi}{4} \pm \arctan\left( \frac{1}{2}e^{-W} \right) = (2n+1)\frac{\pi}{2}, \quad n = 0,1,2... \quad (3.5)$$

For low lying states $e^{-W}$ is usually much smaller than 1 and $\arctan(e^{-W}/2) \cong e^{-W}/2$. Consequently, (3.5) simplifies as,

$$\frac{1}{\hbar}\int_{\theta_0}^{\pi/2} d\theta\sqrt{2J(E - V_0\cos\theta)} \pm \frac{1}{2}e^{-W} = \left( n + \frac{3}{4} \right)\pi. \quad (3.6)$$

However, if in (3.8) we neglect the $e^{-W}/2$ term altogether, we obtain the familiar single-well quantization condition,

$$\frac{1}{\hbar} \int_{\theta_0}^{\pi/2} d\theta \sqrt{2J(E_0 - V_0 \cos\theta)} = \left(n + \frac{3}{4}\right)\pi, \tag{3.7}$$

where we have written $E = E_0$, since in this limit, there is no tunneling splitting. As already mentioned, the tunneling splitting is small for low lying states and high barriers. In this case we can therefore write, $E^{(\pm)} = E_0 \mp \Delta E$, where $\Delta E$ is a small quantity and $E_0$ is determined by (3.7), and

$$\sqrt{2J(E^{(\pm)} - V_0 \cos\theta)} \cong \sqrt{2J(E_0 - V_0 \cos\theta)} \mp \frac{J\Delta E}{\sqrt{2J(E_0 - V_0 \cos\theta)}}.$$

Inserting the above approximation into (3.6), replacing $e^{-W}$ by $e^{-W_0}$, using (3.7) and introducing the frequency of classical motion of the rod in one of the wells,

$$\omega = \frac{2\pi}{T} = \frac{2\pi}{2J \int_{\theta_0}^{\pi/2} d\theta [2J(E_0 - V_0 \cos\theta)]^{-1/2}}, \tag{3.8}$$

results in,

$$E^{(-)} - E^{(+)} = 2\Delta E = \frac{\hbar\omega}{\pi} e^{-W_0}, \tag{3.9}$$

in agreement with (2.7). An alternative derivation of (3.9) is given by Landau and Lifshitz[4], and is further discussed also by Garg[7]. Moreover, for $E_0 \ll V_0$, $\theta = \pi/2 - \theta'$, where $\theta' \ll 1$ and, consequently, $\cos\theta = \sin\theta' \cong \theta'$ and (3.7) simplifies to,

$$E_{0n} = \frac{\hbar^2}{2J} \left(\frac{V_0}{\hbar^2/2J}\right)^{2/3} \left[\frac{3\pi}{2}\left(n + \frac{3}{4}\right)\right]^{2/3}. \tag{3.10}$$

From the derivation given above and from the assumptions contained in WKB approximation, it would be expected that the quantization rules (3.5) and (3.7) are accurate only for $n \gg 1$. However, it turns out that their validity does not depend on the distance between the two

10turning points ($\theta_{0n}$ and $\pi/2$)[5]. We can verify this independently by showing that (3.10) gives reasonable results even for very low lying states, including the ground state, using a different calculation which is outlined in the Appendix.

The WKB wavefunctions well above the barrier ($E \gg V_0$) can be written as in (3.3) with $\theta_0 = -\pi/2$,

$$\psi(\theta) = \frac{C}{\sqrt{L}}\sin\left\{\frac{1}{\hbar}\int_{-\pi/2}^{\theta} d\theta L\right\} = \frac{C}{\sqrt{L}}\cos\left\{\frac{1}{\hbar}\int_{0}^{\theta} d\theta L - \frac{\pi}{4} + \delta\right\}, \quad \delta = \begin{cases} \frac{\pi}{4}, & \psi \text{ even} \\ -\frac{\pi}{4}, & \psi \text{ odd} \end{cases} \quad (3.11)$$

Impossing the boundary conditions, $\psi(\pm\pi/2) = 0$, leads to the quantization condition for the energies $E$,

$$\int_{-\pi/2}^{\pi/2} d\theta L = \hbar n\pi, \tag{3.12}$$

where again, $L = \sqrt{2J(E - V_0\cos\theta)}$, and $n$ is a large integer such that $E \gg V_0$. It is easy to show, using (3.12), that the wavefunctions (3.11) are indeed even or odd with respect to the transformation $\theta \to -\theta$. Of course, in the limit $V_0 \to 0$, (3.11) simplifies to

$$\psi_n(\theta) = \sqrt{\frac{2}{\pi}}\sin n\left(\theta + \frac{\pi}{2}\right), \quad n = 1,2,3,... \tag{3.13}$$

with $E_n = n^2\hbar^2/2J$, corresponding to a »particle« of »mass« $J$ in an infinite potential well of width $\pi$.

When energy $E$ is close to $V_0$, WKB approximation is not applicable in a certain range, $-\theta' < \theta < \theta'$, around the potential maximum[8,9]. In this case we proceed as follows. For sufficiently small values of $\theta$ we can approximate $\cos\theta$ in (2.3) by $1-\theta^2/2$ and, in addition, we assume that this quadratic approximation is valid also for $\theta \gg \theta'$. In this angular range, the Schrödinger equation (2.3) can be written as,

$$\frac{d^2\psi}{d\theta^2} + \frac{JV_0}{\hbar^2}\left[\frac{2(E-V_0)}{V_0} + \theta^2\right]\psi = 0. \tag{3.14}$$

Introducing, $\theta = (\hbar^2/JV_0)^{1/4}\xi = (\hbar/J\omega_c)^{1/2}\xi$, we can rewrite (3.14) in the form,

$$\frac{d^2\psi}{d\xi^2} + (2\varepsilon + \xi^2)\psi = 0, \tag{3.15}$$

where $\varepsilon = (E-V_0)/\hbar\omega_c$. We note, that $(\hbar/J\omega_c)^{1/2}$ and $\hbar\omega_c$ represent natural angular and energy scales, respectively, to measure the angular distance and energy relative to the barrier summit. The differential equation of the type (3.15) occurs when the scalar Helmholtz equation is separated in parabolic coordinates and the solutions are referred to as parabolic cylinder functions[10]. The even and odd solutions will be denoted as $\psi^{(\pm)}(2\varepsilon,\xi)$, and the explicit expressions, which we intend to use, are those given by Morse and Feshbach[8]. As already stated above, we shall assume that the quadratic approximation holds even for very large values of $\xi$ ( for a macroscopic rod $(\hbar/J\omega_c)^{1/2}$ is very small, on the order of $10^{-15}$ in our case), where the asymptotic expressions for $\psi^{(\pm)}(2\varepsilon,\xi)$ may be used. Thus we obtain[8],

$$\psi^{(\pm)}(2\varepsilon,\xi \gg 0) \propto \frac{1}{\sqrt{\xi}}\cos\left\{\frac{1}{2}\xi^2 + \varepsilon\ln\xi - \frac{\pi}{4} \pm \frac{\pi}{8} - \frac{1}{2}\left[(\varphi^{(+)} + \varphi^{(-)}) \pm (\varphi^{(+)} - \varphi^{(-)})\right]\right\}, \tag{3.16}$$

where the phase angles $\varphi^{(\pm)}(2\varepsilon)$ are defined in terms of Gamma functions in the complex plane as,

$$\Gamma\left[\left(\frac{1}{2}\pm\left(-\frac{1}{4}\right)\right) + \frac{1}{2}i\varepsilon\right] = \left|\Gamma\left[\left(\frac{1}{2}\pm\left(-\frac{1}{4}\right)\right) + \frac{1}{2}i\varepsilon\right]\right|e^{i\varphi^{(\pm)}(2\varepsilon)}, \tag{3.17}$$

Using the formulas[10],

$$\Gamma(2z) = (2\pi)^{-1/2}2^{2z-1/2}\Gamma(z)\Gamma(z+1/2),$$

and,

$$\Gamma\left(\frac{1}{4}+iy\right)\Gamma\left(\frac{3}{4}-iy\right)=\frac{\pi\sqrt{2}}{\cosh\pi y+i\sinh\pi y},$$

together with the relation $\Gamma^*(z)=\Gamma(z^*)$, we obtain,

$$\frac{1}{2}\left[\varphi^{(+)}(2\varepsilon)+\varphi^{(-)}(2\varepsilon)\right]=\frac{1}{2}\arg\Gamma\left(\frac{1}{2}+i\varepsilon\right)-\frac{1}{2}\varepsilon\ln 2, \qquad (3.18a)$$

$$\frac{1}{2}\left[\varphi^{(+)}(2\varepsilon)-\varphi^{(-)}(2\varepsilon)\right]=\frac{\pi}{8}-\frac{1}{2}\arctan e^{\pi\varepsilon}. \qquad (3.18b)$$

Inserting this into (3.16), and using the result given by Ford[9] *et al*,

$$\frac{1}{2}\arg\Gamma\left(\frac{1}{2}+i\varepsilon\right)\cong\frac{1}{2}\varepsilon\ln\left[(\varepsilon/e)^2+(1/4\gamma)^2\right]^{1/2}, \quad 4\gamma=4\times 1.78107,$$

yields,

$$\psi^{(\pm)}(2\varepsilon,\xi\gg 0)\propto\frac{1}{\sqrt{\xi}}\cos\left\{\frac{\xi^2}{2}+\frac{\varepsilon}{2}\ln(2\xi^2)-\frac{\varepsilon}{2}\ln\left[(\varepsilon/e)^2+(1/4\gamma)^2\right]^{1/2}\pm\frac{1}{2}\arctan e^{\pi\varepsilon}-\frac{\pi}{4}\right\}.(3.19)$$

Next, we shall assume that in the range of $\xi$ values where the asymptotic expression (3.19) applies, the WKB wavefunctions of the form (3.3) and (3.11), for $\varepsilon<0$ and $\varepsilon>0$, respectively, are also valid. To write the corresponding expressions we must calculate $\int_{\theta_0}^{\theta}d\theta L/\hbar$, using the quadratic approximation for $\cos\theta$. Keeping in mind that $\xi\gg 0$, we obtain,

$$\frac{1}{\hbar}\int_{\theta_0}^{\theta}d\theta L=\int_{\xi_0=\left\{\begin{array}{l}\sqrt{2|\varepsilon|},\varepsilon<0\\ 0,\varepsilon>0\end{array}\right.}^{\xi}d\xi\sqrt{2\varepsilon+\xi^2}\cong\frac{\xi^2}{2}+\frac{\varepsilon}{2}\ln(2\xi^2)-\frac{\varepsilon}{2}\ln(\varepsilon^2/e^2)^{1/2}. \qquad (3.20)$$

Referring to (3.3) and (3.11) and using (3.20), we write the WKB wavefunctions appropriate to this range as,

$$\psi_{WKB}^{(\pm)}(\xi \gg 0) \propto \frac{1}{\sqrt{\xi}} \cos\left\{\frac{\xi^2}{2} + \frac{\varepsilon}{2}\ln(2\xi^2) - \frac{\varepsilon}{2}\ln(\varepsilon^2/e^2)^{1/2} - \frac{\pi}{4} + \delta^{(\pm)} + \begin{pmatrix} 0, \varepsilon < 0 \\ \pm\frac{\pi}{4}, \varepsilon > 0 \end{pmatrix}\right\}. \quad (3.21)$$

The expressions (3.19) and (3.21) must, of course, be identical in the considered range of $\xi$ values. Thus we deduce,

$$\delta^{(\pm)} = \frac{\varepsilon}{2}\ln(\varepsilon^2/e^2)^{1/2} - \frac{\varepsilon}{2}\ln\left[(\varepsilon^2/e^2) + (1/4\gamma)^2\right]^{1/2} \pm \frac{1}{2}\arctan e^{\pi\varepsilon} - \begin{pmatrix} 0, \varepsilon < 0 \\ \pm\frac{\pi}{4}, \varepsilon > 0 \end{pmatrix}. \quad (3.22)$$

Now we can write the WKB wavefunction (we omit the subscript WKB), for energies close to the barrier summit, $E = \varepsilon\hbar\omega_c + V_0$, and for large $\theta$ as,

$$\psi^{(\pm)}(\theta) = \frac{C}{\sqrt{L}}\cos\left\{\frac{1}{\hbar}\int_{\theta_0}^{\theta} d\theta L - \frac{\pi}{4} + \frac{\varepsilon}{2}\ln(\varepsilon^2/e^2)^{1/2} - \frac{\varepsilon}{2}\ln\left[(\varepsilon^2/e^2) + (1/4\gamma)^2\right]^{1/2} \pm \frac{1}{2}\arctan e^{\pi\varepsilon}\right\},$$

(3.23)

where the lower limit in the above integral $\theta_0 = (2\hbar|\varepsilon|/J\omega_c)^{1/2}$ for $\varepsilon < 0$, and $\theta_0 = 0$ for $\varepsilon > 0$. For $\varepsilon < 0$, $e^{\pi\varepsilon} = e^{-W}$, where $W$ is defined after Eq. (3.2). Consequently, for $\varepsilon^2$ large enough we can neglect $(1/4\gamma)^2$ in (3.23) and, if in addition, we replace $\pm\frac{1}{2}\arctan(e^{\pi\varepsilon})$ by $\pm\arctan(e^{\pi\varepsilon}/2)$ for $\varepsilon < 0$, and by $\pm\pi/4$ for $\varepsilon > 0$, we obtain the WKB wavefunctions (3.4) and (3.11), respectively.

Imposing the boundary condition, $\psi^{(\pm)}(\theta = \pi/2) = 0$, gives the quantization condition valid for energies near the barrier summit,

$$\frac{1}{\hbar}\int_{\theta_0}^{\pi/2} d\theta L + \frac{\varepsilon}{2}\ln(\varepsilon^2/e^2)^{1/2} - \frac{\varepsilon}{2}\ln\left[(\varepsilon^2/e^2) + (1/4\gamma)^2\right]^{1/2} \pm \frac{1}{2}\arctan e^{\pi\varepsilon} = \left(n + \frac{3}{4}\right)\pi, \quad (3.24)$$

where $n$ is some large integer. It can be shown[9], using (3.24), that the pairing of even and odd levels into doublets, which is evident well below the barrier (see Eq. 3.9), persists also above the barrier and falls off continuously towards the level structure corresponding to (3.13) in agreement with the numerical calculation presented at the end of section II.

For $\varepsilon = 0$, the even and odd solutions of (3.15) are[8],

$$\psi^{(\pm)}(\varepsilon = 0, \xi > 0) \propto \sqrt{\xi} J_{\mp 1/4}(\xi^2/2), \tag{3.25}$$

where $J_{\mp 1/4}(\xi^2/2)$ are Bessel functions of order ¼. The corresponding asymptotic expressions are[8],

$$\psi^{(\pm)}(\varepsilon = 0, \xi \gg 0) \propto \frac{1}{\sqrt{\xi}} \cos\left\{\frac{1}{2}\xi^2 - \frac{\pi}{4} \pm \frac{\pi}{8}\right\}. \tag{3.26}$$

We notice that the phase difference between even and odd wavefunctions is $\pi/4$, in agreement with (3.16) and (3.22), instead of $\pi/2$ obtained, for large $\varepsilon$, from (3.11).

Finally, we can consider, using WKB wavefunctions, the time evolution (2.15) of the rod with the initial state (2.14). First of all, we can estimate, using the WKB wavefunctions (3.2) and (3.11), that for sufficiently narrow initial state (small $\sigma$) only the energy eigenstates near the barrier summit contribute significantly to the sum (2.15). Consequently, only the wavefunctions (3.23) will be included in (2.15). Moreover, in the limit of macroscopic rod the energy spectrum is quasicontinuous and we can replace the summation by integration with respect to energy. We could calculate the density of states from (3.24), however, to calculate the time of fall of the quantum rod the explicit expression is not needed. Therefore, using (3.23), we write,

$$\psi(\theta \gg 0, t) = \frac{1}{2}\int dE A(E, \theta) \left\{ e^{\frac{i}{\hbar}\left[\int_{\theta_0}^{\theta} d\theta L + \hbar \delta'^{(+)}(\varepsilon) - \frac{\pi \hbar}{4} - Et\right]} + e^{-\frac{i}{\hbar}\left[\int_{\theta_0}^{\theta} d\theta L + \hbar \delta'^{(+)}(\varepsilon) - \frac{\pi \hbar}{4} + Et\right]} \right\}, \tag{3.26}$$

where $E = \hbar\omega_c \varepsilon + V_0$, $A(E, \theta)$ is some real amplitude and $\delta'^{(+)}(\varepsilon)$ is given by (3.22) with the last term omitted. Next, we write

$$\int_{\theta_0}^{\theta} d\theta L = \int_{\theta_0}^{\delta\theta} d\theta L + \int_{\delta\theta}^{\theta} d\theta L = \frac{\hbar}{2}\frac{(\delta\theta)^2}{(\hbar/J\omega_c)} + \frac{\hbar\varepsilon}{2}\ln\left[\frac{2(\delta\theta)^2}{(\hbar/J\omega_c)}\right] - \frac{\hbar\varepsilon}{2}\ln(\varepsilon^2/e^2)^{1/2} + \int_{\delta\theta}^{\theta} d\theta L, \tag{3.27}$$

where $\delta\theta$ is some fixed sufficiently small angle, such that the approximation (3.20) applies. Using this result, the exponentials in (3.26) become,

$$\int_{\theta_0}^{\theta} d\theta L + \hbar \delta^{(+)}(\varepsilon) - \frac{\pi\hbar}{4} \mp Et = \int_{\delta\theta}^{\theta} d\theta L + \frac{\hbar}{2} \frac{(\delta\theta)^2}{(\hbar/J\omega_c)} + \frac{\hbar\varepsilon}{2} \ln\left[\frac{2(\delta\theta)^2}{(\hbar/J\omega_c)}\right] + \hbar \delta''^{(+)}(\varepsilon) - \frac{\pi\hbar}{4} \mp Et,$$

(3.28)

where $\delta''^{(+)}$ is given by (3.20) with the first and last term omitted. To find the motion of the peak of the wave packet (3.26) we use the stationary phase approximation[8]. Equating to zero the derivatives of the exponentials in (3.28) with respect to $E$ we have,

$$J \int_{\delta\theta}^{\theta} \frac{d\theta}{\sqrt{2J(E_0 - V_0 \cos\theta)}} + \frac{1}{2\omega_c} \ln\left[\frac{2(\delta\theta)^2}{(\hbar/J\omega_c)}\right] + \frac{1}{\omega_c} \left.\frac{d\delta''^{(+)}}{d\varepsilon}\right|_{\varepsilon_0} \mp t = 0,$$

(3.29)

where, $E_0 = \hbar\omega_c \varepsilon_0 + V_0$, is the value of $E$ near which the amplitude $A(E,\theta)$ is large. The first term in (3.29) is the time for the classical rod with energy $E_0$ to fall from $\delta\theta$ to $\theta$. Considering the semiclassical limit, we set $E_0 \cong V_0$ ($\varepsilon_0 \cong 0$) and $\theta = \pi/2$, to obtain, using the upper sign in (3.29), the time of fall $t_Q$ for the quantum rod as,

$$t_Q = \frac{1}{\omega_c}\left\{\ln 4(2-\sqrt{2}) - \ln(\hbar/J\omega_c)^{1/2} + \ln(4\gamma)^{1/2} + \pi/4\right\},$$

(3.30)

where the result, $\frac{1}{\omega_c}\int_{\delta\theta}^{\pi/2} \frac{d\theta}{\sqrt{2(1-\cos\theta)}} \cong \frac{1}{\omega_c}\left\{\ln[4(\sqrt{2}-1)] - \ln\delta\theta\right\}$, was also used. We notice that $t_Q$ does not depend on the value of $\delta\theta$ introduced in (3.27) and, moreover, the explicit value for $\sigma$, appart from it being sufficiently small, is not needed. For the rod considered at the end of section II, $t_Q \cong t'_Q \cong 3\sec$.

## IV. THE QUANTUM ROD ON A SLANTED TABLE

Let us assume now that the table top is not absolutely horizontal but is slightly slanted at an angle $\delta\theta \ll 1$ with respect to the horizontal plane. The time-independent Schrödinger equation (2.3) remains the same only the boundary conditions are changed to $\psi\left(\pm\frac{\pi}{2} - \delta\theta\right) = 0$. However, if we introduce a new variable $\theta' = \theta + \delta\theta$, we can write the Hamiltonian for the rod on the slanted table as, $H' = H + V'$ where,

$$H = -\frac{\hbar^2}{2J}\frac{d^2}{d\theta'^2} + V_0' \cos\theta', \tag{4.1}$$

$$V' = V_0 \sin\delta\theta \sin\theta' \cong V_0 \delta\theta \sin\theta', \tag{4.2}$$

and $V_0' = V_0 \cos\delta\theta \cong V_0$. The eigenfunctions of (4.1), $H\psi(\theta') = E\psi(\theta')$, subject to the boundary conditions $\psi(\theta' = \pm\pi/2)$, and the corresponding eigenvalues are the same as the eigenfunctions and the eigenvalues discussed above except that everywhere $V_0$ must be replaced by $V_0'$. Since $\delta\theta \ll 1$, the eigenfunctions $\psi_n'$ and eigenvalues $E_n'$ of $H'$ are determined by considering $V'$ as a small perturbation. In particular, we are interested here only in the eigenfunctions $\psi_n'$ corresponding to the energies $E_n' \ll V_0'$. Let us choose $\delta\theta$ such that,

$$\left[\left(\frac{E_{n+1}'^{(-)} + E_{n+1}'^{(+)}}{2}\right) - \left(\frac{E_n'^{(-)} + E_n'^{(+)}}{2}\right)\right] \gg V_0 \delta\theta \gg E_n'^{(-)} - E_n'^{(+)}. \tag{4.3}$$

In this case we can solve the eigenvalue problem,

$$H'\psi_n' = E_n'\psi_n', \tag{4.4}$$

for each doublet ($\psi_n'^{(\pm)}$) separately (we have essentially a two level system). Thus we write, omitting the subscript $n$ in what follows,

$$\psi' = c^{(+)}\psi'^{(+)} + c^{(-)}\psi'^{(-)}, \tag{4.5}$$

Inserting this into (4.4), using $H\psi'^{(\pm)} = E'^{(\pm)}\psi'^{(\pm)}$, multiplying the resulting equation once by $\psi'^{(+)}$ and then by $\psi'^{(-)}$ and integrating each time over $\theta'$ from $-\pi/2$ to $+\pi/2$, yields,

$$c^{(+)}\left(E'^{(+)} - E'\right) + c^{(-)}V_{(\pm)}' = 0, \tag{4.6a}$$

$$c^{(+)}V_{(\mp)}' + c^{(-)}\left(E'^{(+)} - E'\right) = 0, \tag{4.6b}$$

where,

$$V'_{(\pm)} = V'_{(\mp)} \equiv V_0 \delta\theta \int_{-\pi/2}^{\pi/2} d\theta' \sin\theta' \psi'^{(-)}(\theta')\psi'^{(+)}(\theta').  \quad (4.7)$$

The solution of (4.6) is,

$$E' = \frac{1}{2}\left(E'^{(-)} + E'^{(+)}\right) \pm \frac{1}{2}\sqrt{\left(E'^{(-)} - E'^{(+)}\right)^2 + 4V'^2_{(\pm)}}, \quad (4.8)$$

and

$$\psi' = c'\left\{\psi'^{(+)} \pm \psi'^{(-)}\sqrt{1 + \frac{1}{4}\frac{\left(E'^{(-)} - E'^{(+)}\right)^2}{V'^2_{(\pm)}}} + \psi'^{(-)}\frac{1}{2}\frac{\left(E'^{(-)} - E'^{(+)}\right)}{V'_{(\pm)}}\right\}. \quad (4.9)$$

In the classical limit, i.e. with macroscopic rods, the conditions (4.3) can be easily satisfied. Moreover, for macroscopic rods $(E'^{(-)} - E'^{(+)}) \propto e^{-W_0} \to 0$ and (4.9) becomes

$$\psi' = \frac{1}{\sqrt{2}}\left(\psi'^{(+)} \pm \psi'^{(-)}\right). \quad (4.10)$$

Recalling (2.4), we conclude, that in this limit the eigenstates well below the top of the barrier are, contrary to the case of absolutely horizontal table, localized in one or the other well, no matter how small is $\delta\theta$.[11]

## V. CONCLUSIONS

A rigid rod with one end fixed to a horizontal table has been analyzed in detail using WKB approximation. In particular, the WKB wavefunctions near the barrier top were determined and were used to calculate the time of fall of the quantum rod. The result obtained confirms the qualitative statement made previously by Tegmark and Wheeler[1]. As already pointed out, the first term in (3.29) stands for the time for classical rod to fall from $\delta\theta$ to $\theta$. The remaining

4terms are then, obviously, the time needed for the rod to fall from the upright orientation, represented by the initial state (2.14), to $\delta\theta$. It has been implicitly assumed in this analysis, that each half of the wave packet describing the quantum rod falling symmetrically in both directions, holds together without too much spreading for a time long compared to the time of fall. The characteristic time at which a quantum wave packet representing the rod begins to spread out significantly is[12],

$$t_0 = \frac{2J\sigma^2}{\hbar}, \qquad (5.1)$$

where $\sigma$ is the width of the initial state. If we write, $\sigma = \alpha(\hbar/J\omega_c)^{1/2}$, we obtain $t_0 = 2\alpha^2/\omega_c$. For a 10 cm long rod with mass equal to $10^{-3}$kg $\omega_c \approx 10\text{s}^{-1}$, consequently, $\alpha^2 >> 10$. If we take $\alpha = 10$, we still have a very narrow initial state.

In addition, it has been also shown that the rod balanced initially on a slightly slanted table, will not fall down symmetrically because, in this case, the ground state is localized in the lower well. This instability of tunneling against small potential asymmetries has been pointed out some time ago by Claverie and Jona-Lasinio[11] and was rediscovered recently by Gea-Banacloche[13]. The same phenomenon was discussed, in a different context, also by the authors[14]. To conclude, let us mention that using time-dependent Schrödinger equation corresponding to (2.2) a numerical simulation of the falling rod can be performed rather easy. However, the rod does not just fall, it bounces back and falls again et cetera, clearly accompanied by the wave packet spreading if the initial state is too narrow.

**APPENDIX**

Low lying energy eigenstates are confined to the bottom of the wells and, consequently $|\theta| = \pi/2 - \theta'$, where $\theta' << 1$. Therefore we can rewrite the time-independent Schrödinger equation (2.3), corresponding to the wavefunctions $\psi_0$ localized in the wells, as

$$-\frac{d^2\psi_0}{d\theta'^2} + \left(\frac{V_0}{\hbar^2/2J}\theta' - \frac{E_0}{\hbar^2/2J}\right)\psi_0 = 0, \qquad (A1)$$

with the boundary condition $\psi_0(\theta' = 0) = 0$. (A1) represents motion in a linear potential with the classical turning point $\theta'_0 = E_0/V_0$ (the other turning point is, of course, at $\theta' = 0$). If we introduce a new variable $\eta = \left(\dfrac{V_0}{\hbar^2/2J}\right)^{1/3}\theta' - \lambda$, where

$$\lambda = \frac{E_0}{\hbar^2/2J} \cdot \left(\frac{\hbar^2/2J}{V_0}\right)^{2/3} \tag{A1}$$

becomes,

$$\frac{d^2\psi_0}{d\eta^2} - \eta\psi_0 = 0, \tag{A2}$$

and the boundary condition is now $\psi_0(-\lambda) = 0$. The classical motion is therefore restricted to $-\lambda \leq \eta \leq 0$. The solutions of (A2), which are finite for $\eta \to \infty$ or $\theta' \gg \theta'_0$, are Airy functions[4,15]. The energy eigenvalues are determined from the boundary condition $\psi_0(-\lambda_n) = 0$, where $-\lambda_n$ is recognized as the $n$-th zero of the Airy function[10]. Thus we have

$$E_{0n} = \frac{\hbar^2}{2J} \cdot \left(\frac{V_0}{\hbar^2/2J}\right)^{2/3} \lambda_n. \tag{A3}$$

Comparing (A3) and (3.10) we can write, $\lambda_n(WKB) = \left[\dfrac{3\pi}{2}\left(n + \dfrac{3}{4}\right)\right]^{2/3}$. A few numerical values are given in Table 2.

21**Tables**

**Table 1**: Energy eigenvalues (2.13) obtained by numerical solution of Mathieu equation (2.8), corresponding to $V_0/(\hbar^2/2J) = 10^4$. The numbers in the last column represent the measure of the pairing effect.

| n | $E_n^+/(\hbar^2/2J)$ | $E_n^-/(\hbar^2/2J)$ | $\dfrac{E_n^- - E_n^+}{(\hbar^2/2J)}$ | $\dfrac{E_{n+1}^+ - E_n^+}{(\hbar^2/2J)}$ | $\dfrac{E_n^- - E_n^+}{E_{n+1}^+ - E_n^+}$ |
|---|---|---|---|---|---|
| 23 | 9420.43 | 9420.43 | 0.00 | 187.59 | 0.0000 |
| 24 | 9608.02 | 9608.03 | 0.01 | 170.31 | 0.0001 |
| 25 | 9778.33 | 9778.70 | 0.37 | 143.94 | 0.0026 |
| 26 | 9922.26 | 9930.30 | 8.04 | 102.02 | 0.0788 |
| 27 | 10024.28 | 10071.29 | 47.01 | 122.49 | 0.3838 |
| 28 | 10146.77 | 10223.21 | 76.45 | 159.71 | 0.4786 |
| 29 | 10306.48 | 10394.20 | 87.72 | 179.67 | 0.4882 |
| 30 | 10486.15 | 10581.94 | 95.79 | 195.17 | 0.4908 |
| 31 | 10681.32 | 10784.10 | 102.78 | 208.79 | 0.4923 |
| 32 | 10890.11 | 10999.23 | 109.12 | 221.23 | 0.4932 |
| 33 | 11111.34 | 11226.37 | 115.02 | 232.87 | 0.4939 |
| 34 | 11344.21 | 11464.82 | 120.61 | 243.92 | 0.4945 |

**Table 2**: When tunneling is neglected the lowest eigenvalues of (2.3) are calculated using WKB approximation (3.10) or, in terms of Airy functions, as solutions of (A2). A few lowest energies, expressed in units of $\left(\dfrac{\hbar^2}{2J}\right)\left(\dfrac{V_0}{\hbar^2/2J}\right)^{2/3}$, obtained by WKB or by exact solutions for the linear potential, are presented in columns 2 and 3, respectively, in Table 2.

| n | $\lambda_n$(WKB) | $\lambda_n$ |
|---|---|---|
| 0 | 2.320 | 2.338 |
| 1 | 4.082 | 4.088 |
| 2 | 5.517 | 5.521 |
| 3 | 6.784 | 6.787 |
| 4 | 7.942 | 7.944 |
| 5 | 9.021 | 9.023 |

**Figures:**

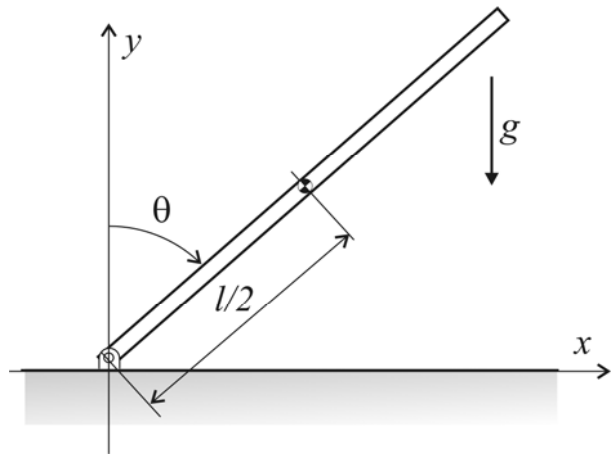

**Fig.1**: Rigid rod with one end fixed on a horizontal table.

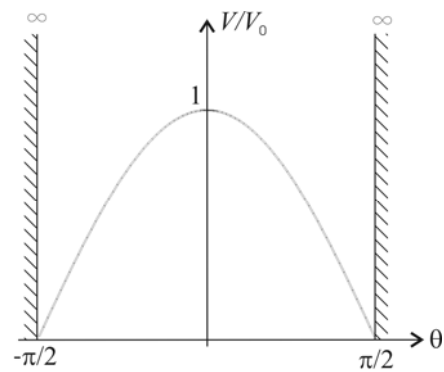

**Fig.2**: Schematic representation of the gravitational potential energy, $V = V_0\cos\theta$, of a rigid rod with one end fixed on a horizontal table.

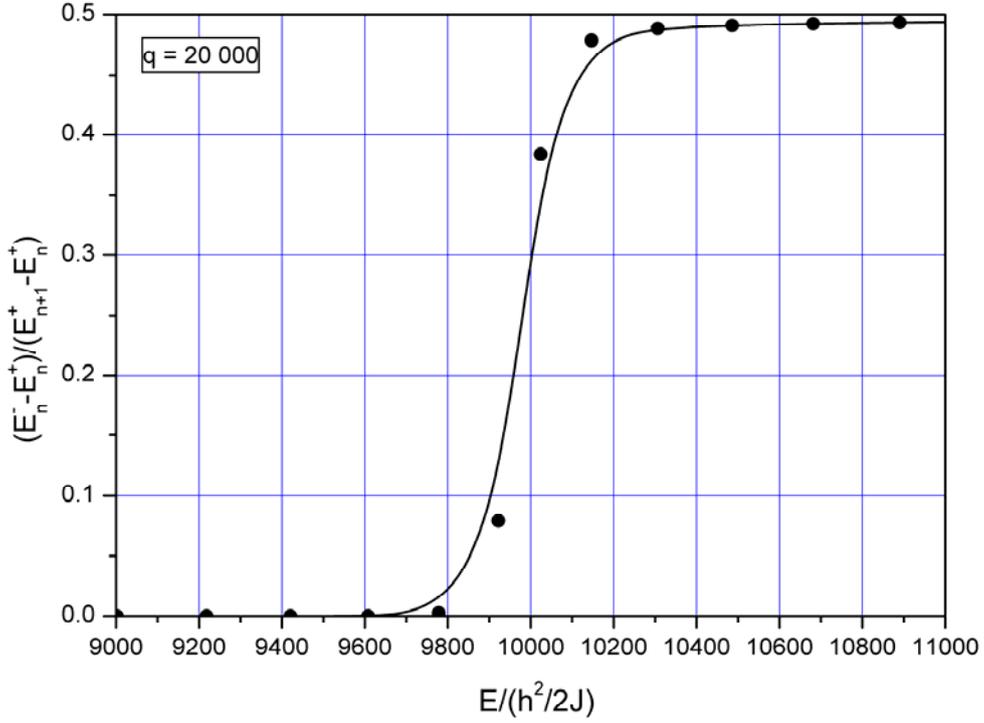

**Fig.3**: The measure of the pairing effect, expressed as a ratio of the energy difference between an odd level and the preceding even level, and the energy difference between two successive even levels. The energy values on the horizontal axis are given in units of ($\hbar^2/2J$). The corresponding numerical values are listed in the last column of Table 1 for $V_0/(\hbar^2/2J) = 10^4$. In the limit $E\to\infty$, the ratio approaches the value ½, appropriate for the energy spacings of a particle in an infinite potential.